\documentclass[11pt]{article}
\usepackage{latexsym}
\usepackage{epsf}
\usepackage{cite}

\def\be{\begin{equation}}
\def\ee{\end{equation}}
\def\bea{\begin{eqnarray}}
\def\eea{\end{eqnarray}}
\def\ba{\begin{array}}
\def\ea{\end{array}}

\begin{document}
\begin{flushright}
IFUM-762-FT\\
\end{flushright}
\vspace{1truecm}

\centerline{\Huge Kinky D-branes and straight strings}

\centerline{\Huge of open string tachyon effective theory}

\vspace{2truecm}

\begin{center}
 {\Large  Luca~Martucci\footnote{luca.martucci@mi.infn.it}
 and Pedro~J.~Silva\footnote{pedro.silva@mi.infn.it}}\\
\renewcommand{\thefootnote}{\arabic{footnote}}
\vspace{.5truecm}
Dipartimento di Fisica dell'Universit\`a di Milano,\\
Via Celoria 16, I-20133 Milano, Italy\\

\vspace*{0.5cm}

INFN, Sezione di Milano,\\
Via Celoria 16,
I-20133 Milano, Italy\\

\end{center}

\vspace{2truecm}

%%%%%%%%%%%%%%%%%%%%%%%%%%%%%%%%%%%%%%%%%%%%%%%%%%%%%%
\centerline{\bf ABSTRACT}
\vspace{1truecm}
\noindent
In this letter we construct the kink D1-brane super D-helix solution and its T-dual the D2-brane supertube using the effective action of non-BPS tachyonic D-branes . In the limit of zero angular momentum, both types of solutions collapse to zero radius, giving rise respectively to a degenerate string configuration corresponding to a particle travelling with the speed of light and to a static straight string configuration. These solutions share all the properties of fundamental strings and do not have the pathological behavior of other solutions previously found in this context. A short discussion on the ``generalized gauge transformations'' suggested by Sen is used to justify the validity of our approach.
\vspace{.5truecm}

%%%%%%%%%%%%%%%%%%%%%%%%%%%%%%%%%%%%%%%%%%%%%%%%%%%%%%%%%
\newpage

\noindent {\Large \em Introduction}
\vspace{.5truecm}

Open string tachyons are nowadays seen with better eyes than ever. Ten years ago the presence of a tachyon was basically a signal of  failure. Nevertheless in the last few years a lot of effort has been devoted to understand the physics that tachyon can unravel, at least in the context of string theory. In fact tachyons seems to define useful frameworks where to study open/closed string theory relations and in particular non-BPS D-brane physics \cite{9909062,0003122,0003221,0004106,0204143,0209122,0301076,0303035,0303139,0304045,0304108,0011226,0012222,0208217,0301101,0301179,0304197,0303057,0012080,0104218,0202079,0208142,0304180,0301049}.

By now, it is well established that BPS D-brane physics can be studied within the open string tachyon framework. In particular in \cite{0303057}, it is shown that any D-brane configuration can be reproduced using tachyonic kink solutions, in the sense that the D-brane and its fluctuations are identified with the kink and its corresponding fluctuations. On the other hand, it would be very interesting to study other extended objects appearing in M-theory, like for example NS-branes and fundamental strings. Regarding the strings, in \cite{0305011} a detailed analysis of the characteristics that a tachyonic solution describing fundamental strings should have is given and it is shown that previous candidates like those of \cite{0002223,0009061,0010240,9901159,0209034,0305092,LIND} should not be identified with fundamental strings. In that work, as a result of the general properties of the kink D-brane solutions discussed in \cite{0303057}, a new family of kink configurations is presented and is argued that they correspond to fundamental strings attached to D-branes. Nevertheless, finding other D-brane configurations with magnetic flux (to describe for example non-commutative theories) may proved to be more involved, since these fluxes in turn could induce topological obstructions to continue the gauge field outside the kink. 

Returning to the fundamental strings, if we believe in the ``p-brane democracy principle'' \cite{9507048}, at least some of the properties of the strings should be accessible to us using already understood tools borrowed from D-brane physics. In fact, there are frameworks where fundamental strings are described in terms of D-branes degrees of freedom, like matrix string theory \cite{mst} or the BFSS conjecture \cite{bfss}, and more recently using non-abelian D-brane effective actions \cite{9910053,9910052}. Actually, there are a couple of very good example of D-brane configurations that in a particular limit leave us with only string states, first the so-called supertube \cite{0103030,0106012,0112054,0112034,0205265}, a bound state of many D0-branes and fundamental strings, that due to the presence of angular momentum blows up to stable D2-branes with electric and magnetic fields dilute in its worldvolume, and second the D-helix \cite{0105095}, a bound state of D1-branes and fundamental strings that again due to angular momentum acquires the shape of a helix. These two configuration are related by T-duality and in the limit of zero angular momentum reduce to  type IIA  straight string and type IIB strings collapsed to a point traveling in a straight line at the speed of light. 

In this letter we make notice that, using the ``generalized gauge principle'' suggested by Sen in \cite{0303057}, the above type of D-brane configurations can be described using a tachyonic kink solution of non-BPS D-branes action, where the possible topology obstruction can be somehow by-passed leaving the correct regular effective theory on the kink D-brane. This ``generalized gauge principle'' is related to the fact that the stress-energy tensor of the kink solutions is zero everywhere but on the kink itself, rendering unphysical any fluctuation outside the kink. Therefore, from the point of view of the theory on the kink, the shape of the continued gauge field is identified with gauge degrees of freedom and hence is irrelevant for the corresponding D-brane phenomenology. Also we would like to emphasize that we consider the non-BPS D-brane action of \cite{0303057} as an effective description in the sense that gives the right physics only in an appropriated regime and is in that respect that we work on this letter. 

In the next section we construct the  D-helix and then the supertube, both as tachyonic kink solutions of unstable non-BPS D-branes. Then we discuss on the validity of the solutions and the role of the  ``generalized gauge principle''. At the end we take the limit of zero angular momentum to obtain the kink superstrings of type IIA and type IIB.

\vspace{1truecm}
\noindent {\Large \em The kinky D-helix and supertube}  
\vspace{.5truecm}

Let us briefly summarized the D-helix solution of \cite{0105095}. It is convenient to write the ten dimensional flat metric as
\bea
\label{met}
ds^2_{(10)}=-dt^2+dX^2+R^2d\phi+dR^2+ds^2(E^2)\ ,
\eea
where cylindrical coordinates are used in the first four directions. Then, the D-helix solution is a D1-brane described by the following embedding 
\bea
&&t=\tau\quad,\quad\phi=\sigma\ (\hbox{mod}\ 2\pi)\ ,\cr
&&R=R_0\quad,\quad X=\tau E-\sigma B\ ,
\eea
where all the other space-time coordinates are fixed and the worldsheet coordinates ($\tau,\sigma$) are defined with range $(-\infty,+\infty)$. 
The Born-Infeld Lagrangian for these configurations is given by 
\bea
{\cal L}^{(D1)}=-T_{D1}\sqrt{(1-E^2)R_0^2+B^2}\ 
\eea
while the Hamiltonian is
\bea
{\cal H}^{(D1)}=\sqrt{(\pi_X\pm T_{D1}B)^2+(\hbox{${\pi_X B\over R_0}$}\mp T_{D1}R_0)^2}
\eea
where $\pi_X=T_{D1}R_0^2E/\sqrt{(1-E^2)R_0^2+B^2}$ is the conjugated momentum of $X$. The Hamiltonian analysis just tell us that $\pi_X$ is conserved, being related to the number of strings diluted on the D1-brane. On these solutions, given $\pi_X$ and $B$, the Hamiltonian is bounded from below and acquires its minimal value only when $\pi_XB=\pm T_{D1}R_0^2$ that implies $E^2=1$ corresponding to the supersymmetric solution preserving $1/4$ of the maximal supersymmetry.  

Following the general prescription given in \cite{0303057}, we can recover the D-helix as a kink solution of a non-BPS D2-brane. The non-BPS D2-Brane effective action is 
\be
\label{uda}
S_{D2}=-\int d\xi^{3}\ V(T)\sqrt{-det(g_{ij}+\partial_i T\partial_j T +F_{ij})}
\ee
where $g_{ij}$ the pull-back of the metric with $i=(0,1,2)$, labeling the worldvolume directions and $F_{ij}$ the worldvolume abelian Yang-Mills (YM) field strength and $T$ is the opens string tachyon with potential $V(T)$. This potential is taken to be even in $T$, with maximum at $T=0$, equal to the tension of the non-BPS D2-brane, and minimum at $T=\pm\infty$ where it vanishes.

We take the worldvolume coordinates on the D2-brane to be $\xi^i=(\tau,\sigma,\rho)$ where $(\tau, \sigma)$ have the same range as before and $\rho$ runs in the interval $[0,\infty)$. Using the same metric as in (\ref{met}) we consider the ansatz\footnote{Given the identification of $\phi$ with $\sigma$ modulo $2\pi$, the above embedding could be seen as a static gauge using different coordinates patches.}
\bea
\label{ansatz1}
&&t=\tau\quad,\quad\phi=\sigma\ (\hbox{mod}\ 2\pi)\cr 
&&R=\rho\quad,\quad X=\Omega(\rho)(\tau E-\sigma B),
\eea
with $F_{ij}=0$ and where $\Omega(\rho)$ is a regular function different from zero only in a small neighborhood of $R_0$ with $\Omega(R_0)=1$ and $\Omega^\prime(R_0)=0$. For the tachyon field we use
\be
\label{tac2}
T=f(a\ln\hbox{${\rho\over R_0}$}).
\ee
where $f(u)$ is odd in $u$ and $f^\prime(u)> 0$ for all $u$. Therefore in the limit $a\rightarrow \infty$ we have
\bea\label{tac}
T(\rho)=\left\{\begin{array}{cc}
            \infty \quad \hbox{if} \quad \rho>R_0\\
            0 \quad \hbox{if} \quad \rho=R_0\\
	    -\infty \quad \hbox{if} \quad \rho<R_0
            \end{array}\right. 
\eea
Substituting this ansatz in the action we obtain that
\bea
\label{uda2}
{\cal L}^{(D2)}= V(f)\sqrt{1+\hbox{($\frac{ af^\prime}{\rho})^2$}}\sqrt{(1-\Omega^2 E^2)\rho^2+\Omega^2B^2+O(\Omega^\prime)}\ .
\eea
Then the conjugate momentum of $X(\xi)$ is given by
\bea
\Pi_X = \frac{\partial {\cal L}^{(D2)}}{\partial(\partial_\tau X)}= V(f)\sqrt{1+\hbox{($\frac{ af^\prime}{\rho})^2$}}\left[ \frac{\rho^2\Omega E}{\sqrt{(1-\Omega^2E^2)\rho^2+\Omega^2B^2}}+O(\Omega^\prime)\right] \ ,
\eea
and the Hamiltonian density takes the form
\bea
{\cal H}^{(D2)}= \frac 1\rho   V(f)\sqrt{1+\hbox{($\frac{ af^\prime}{\rho})^2$}}\sqrt{\big[ \rho^2 +\Omega^2B^2\big]\big[\rho^2+\hbox{$\frac{\Pi_X^2}{ V(f)^2\left(1+\hbox{$(\frac{af^\prime}{\rho})^2$}\right)}$})\big](1+O(\Omega^\prime))}\ .
\eea
Then using the following normalization on the tachyon potential $V(T)$
\bea
\int_0^{+\infty} d\rho \frac a\rho f^\prime V(f)=\int_{-\infty}^{+\infty} df V(f)=T_{D1}\ ,
\eea
for any function $A(\rho)$, it is verified that 
\bea
&&\lim_{a\rightarrow \infty} \int_{o}^\infty  d\rho \frac a\rho f^\prime  V(f) \; A(\rho) = \cr 
&&=\int_{-\infty}^\infty  duV(u)A(R_0exp{(\hbox{${f^{-1}(u)/a}$})})=A(R_0)T_{D1}\ .
\eea 
Hence, using the fact that $\Omega^\prime(R_0)=0$, we obtain the following relations: 
\bea
&&\lim_ {a \rightarrow \infty}  \int d\rho\ {\cal L}^{(D2)}= {\cal L}^{(D1)}\, \cr
&&\lim_ {a \rightarrow \infty} \int d\rho\ \Pi_X = \pi_X \ ,\cr
&&\lim_ {a \rightarrow \infty} \int d\rho \ {\cal H}^{(D2)} = {\cal H}^{(D1)}\ .
\eea
We then see that, for fixed $\pi_X$ and $B$ we recover that the Hamiltonian is minimized if $T_{D1}R_0=|\pi_X B|$.

Therefore, in accordance with the general analysis of  \cite{0303057}, we obtain a kink solution corresponding to the D-helix presented above with the same conserved charges and the same supersymmetric bound. Furthermore, from \cite{0303057} we also know that even the complete dynamics of the fluctuations around the D-helix configuration is recovered as the dynamics of the fluctuations around the kink solution (\ref{ansatz1},\ref{tac2}).

Let us consider now the supertube. These configurations are related to the D-helix by a T-duality transformation along the $X$-direction. For the circular D-helix, the result is a D2-brane with a cylindric shape of radius $R_0$, with  worldvolume coordinates (written in the static gauge) $(\tau,\sigma,x)$  that have the following ranges: $\sigma \in [0,2\pi)$ and $(\tau,x) \in (-\infty,+\infty)$. Also, there is a non-zero electromagnetic field of the form
\bea
F=Ed\tau\wedge dx + B dx\wedge d\sigma\ ,
\eea
that stabilizes the brane preventing its collapse to zero radius. As in the D-helix case, there is a conserved momentum (this time conjugated to the gauge field $A_x$) $\pi_{A_x}$, that once fixed, together with $B$ parametrizes the solutions. In particular, for $R_0^2=B\pi_{A_x} $ the minimal energy supersymmetric configuration is obtained, setting $E^2=1$. This last configuration preserves the same amount of supersymmetries as the supersymmetric D-helix. 

To describe the supertube in terms of a tachyonic kink, we use the action of a non-BPS D3-brane, 
\be
%\label{uda}
S_{D3}=-\int d\xi^{4}\ V(T)\sqrt{-det(g_{ij}+\partial_i T\partial_j T +F_{ij})}\ ,
\ee
with worldvolume coordinates $(\tau,x,\sigma,\rho)$ where $(\tau,x,\sigma)$ have the same range as before and $\rho$ runs on the interval $[0,\infty)$. We then use an ansatz inspired by the D-helix,
\bea
\label{ansatz2}
t&=&\tau\quad,\quad\phi=\sigma\ ,\cr 
R&=&\rho\quad,\quad X=x\ , \cr
A_x&=&\Omega(\rho)(\tau E-\sigma B)\ ,
\eea
where all the other fields are set to zero and $\Omega(\rho)$ has the same properties as in (\ref{ansatz1}), the metric is like in (\ref{met}) and  $T(\rho)$ is set to be of the same form as in (\ref{tac2}). 

Note that in principle there are problems with the gauge field $A_x$ defined in (\ref{ansatz2}). First of all, for the supertube $\sigma$ is identified modulo $2\pi$ while $\delta_{rot} A_x=A_x(\sigma +2\pi)-A_x(\sigma)=-2\pi B\Omega(\rho)$. This change in $A_x$ can not be reabsorbed by an ordinary gauge transformation since $d(\delta_{rot} A)\neq 0$ due to the nontrivial dependence on $\rho$. Nevertheless, Sen introduced in \cite{0303057} a ``generalized gauge equivalence principle'' that identifies kink configurations with different profiles of the non-BPS worldvolume fields outside the location of the associated BPS brane i.e. outside the kink itself. Therefore we can perform an ordinary gauge transformation $\delta_\Lambda A= d\Lambda$ such that $(\delta_{rot}+\delta_\Lambda)A_x= 0$ and $\delta_\Lambda A_\rho\propto \Omega^\prime$ and then reinterpret the new $A_\rho$ as a pure generalized gauge degree of freedom, since $\Omega^\prime(R_0)=0$\footnote{We could write the configuration (\ref{ansatz2}) using two patches where on each one the fields are well defined and on the overlapping region they are related by a generalized gauge transformation.}. Second, the above configurations come with a topological obstruction, that can be seen as a monopole source inside the tube i.e. for $\rho<R_0$. But, we should stress that the field strength $F$ of the non-BPS branes is invariant under the generalized gauge transformation only on the kink and therefore the usual topological obstruction becomes unphysical since we can change the flux outside the kink by reshaping the field strength. In fact we can even make it zero for a limiting case like the following $A_x=E\tau , A_\sigma=\Omega(\rho)xB$, where the gauge field is well defined everywhere. In this case we have a non-zero $F_{\rho\sigma}=x\Omega^\prime(\rho)B$ that is a pure generalized gauge for every $x$. 

Then, using the ansatz (\ref{ansatz2}) we  recover the supertube as a tachyonic kink. Details of the calculation are so similar to the D-helix case that for brevity they are repeated here but, from the general discussion of \cite{0303057}, we known that not only the exact charges and supersymmetric bounds are recovered, but also the dynamics of its fluctuation. 

\vspace{1truecm}
\noindent {\Large \em Straight Strings}
\vspace{.5truecm}

As a last comment, we can always take the particular limit on the above solutions (D-helix and supertube) of vanishing $B$ keeping constant the conserved momentum $\pi_X$ or $\pi_{A_x}$ respectively. In this way the radius $R_0$ is sent to zero keeping the regularity of the solution. These limiting configurations correspond to a degenerate superstring vacuum state or superparticle in type IIB traveling in the $X$-direction with the velocity of light, and to a straight superstring vacuum in type IIA along the $X$-direction. These configurations are well defined as kink solutions having the correct tension and Hamiltonian density characteristic of fundamental strings and in the type IIA case the worldvolume is exactly two dimensional while the electric flux is confined in a one-dimensional region (the $X$-direction). All the properties expected to hold for a fundamental string are attained\footnote{See \cite{0305011} for a detail explanation on this point.} since the tachyon field is infinite outside the worldvolume of the string but is finite inside. Therefore these solutions correspond to a realization of the program developed by Sen in the case of a single string\footnote{In principle the solution can represent more than one string, depending on the value of conserved momentum.} unattached to any D-brane. Also the dynamics of the perturbation will be described in terms of a supersymmetric two dimensional Nambu-Goto action in both cases.

We would like to stress that although the solutions here found have cylindrical shape, it is trivial to extend the analysis to the case of general shape in curved backgrounds like in \cite{0304210} and we have used such a restrictive ansatz for the shake of simplicity.

%%%%%%%%%%%%%%%%%%%%%%%%%%%%%%%%%%%%%%%%%%%%%%%%%%%%%%%%%%%%%%%%%%%%%%%%%%%%%%%%
\vspace{.5cm}
{\bf Acknowledgments}\\

We thank S. Cacciatori, M. Caldarelli, A. Celi,  D. Mateos and J. Simon for useful discussions. This work was supported by INFN, MURST and by the European Commission RTN program HPRN-CT-2000-00131, in association with the University of Torino.

%%%%%%%%%%%%%%%%%%%%%%%%%%%%%%%%%%%%%%%%%%%%%%%%%%%%%%%%%%%%%%%%%%%%%%%%%%%%%%%

%%%%%%%%%%%%%%%%%%%%%%%%%%%%%%%

\vspace{1cm}

\begin{thebibliography}{99}

\bibitem{9909062}
A.~Sen,
%``Supersymmetric world-volume action for non-BPS D-branes,''
JHEP {\bf 9910}, 008 (1999)
[arXiv:hep-th/9909062].
%%CITATION = HEP-TH 9909062;%%

\bibitem{0003122}
M.~R.~Garousi,
%``Tachyon couplings on non-BPS D-branes and Dirac-Born-Infeld action,''
Nucl.\ Phys.\ B {\bf 584}, 284 (2000)
[arXiv:hep-th/0003122];
%%CITATION = HEP-TH 0003122;%%
%``On-shell S-matrix and tachyonic effective actions,''
Nucl.\ Phys.\ B {\bf 647}, 117 (2002)
[arXiv:hep-th/0209068];
%%CITATION = HEP-TH 0209068;%%
arXiv:hep-th/0303239;
%%CITATION = HEP-TH 0303239;%%
arXiv:hep-th/0304145.
%%CITATION = HEP-TH 0304145;%%

\bibitem{0003221}
E.~A.~Bergshoeff, M.~de Roo, T.~C.~de Wit, E.~Eyras and S.~Panda,
%``T-duality and actions for non-BPS D-branes,''
JHEP {\bf 0005}, 009 (2000)
[arXiv:hep-th/0003221].
%%CITATION = HEP-TH 0003221;%%

\bibitem{0004106}
J.~Kluson,
%``Proposal for non-BPS D-brane action,''
Phys.\ Rev.\ D {\bf 62}, 126003 (2000)
[arXiv:hep-th/0004106].
%%CITATION = HEP-TH 0004106;%%

\bibitem{0204143}
A.~Sen,
%``Field theory of tachyon matter,''
Mod.\ Phys.\ Lett.\ A {\bf 17}, 1797 (2002)
[arXiv:hep-th/0204143].
%%CITATION = HEP-TH 0204143;%%

\bibitem{0209122}
A.~Sen,
%``Time and tachyon,''
arXiv:hep-th/0209122.
%%CITATION = HEP-TH 0209122;%%

\bibitem{0301076}
C.~j.~Kim, H.~B.~Kim, Y.~b.~Kim and O.~K.~Kwon,
%``Electromagnetic string fluid in rolling tachyon,''
arXiv:hep-th/0301076.
%%CITATION = HEP-TH 0301076;%%

\bibitem{0303035}
F.~Leblond and A.~W.~Peet,
%``SD-brane gravity fields and rolling tachyons,''
arXiv:hep-th/0303035.
%%CITATION = HEP-TH 0303035;%%

\bibitem{0303139}
N.~Lambert, H.~Liu and J.~Maldacena,
%``Closed strings from decaying D-branes,''
arXiv:hep-th/0303139.
%%CITATION = HEP-TH 0303139;%%

\bibitem{0304045}
D.~Kutasov and V.~Niarchos,
%``Tachyon effective actions in open string theory,''
arXiv:hep-th/0304045.
%%CITATION = HEP-TH 0304045;%%

\bibitem{0304108}
K.~Okuyama,
%``Wess-Zumino Term in Tachyon Effective Action,''
arXiv:hep-th/0304108.
%%CITATION = HEP-TH 0304108;%%

\bibitem{0011226}
J.~A.~Minahan and B.~Zwiebach,
%``Gauge fields and fermions in tachyon effective field theories,''
JHEP {\bf 0102}, 034 (2001)
[arXiv:hep-th/0011226].
%%CITATION = HEP-TH 0011226;%%

\bibitem{0012222}
M.~Alishahiha, H.~Ita and Y.~Oz,
%``On superconnections and the tachyon effective action,''
Phys.\ Lett.\ B {\bf 503} (2001) 181
[arXiv:hep-th/0012222].
%%CITATION = HEP-TH 0012222;%%

\bibitem{0208217}
N.~D.~Lambert and I.~Sachs,
%``Tachyon dynamics and the effective action approximation,''
Phys.\ Rev.\ D {\bf 67}, 026005 (2003)
[arXiv:hep-th/0208217].
%%CITATION = HEP-TH 0208217;%%

\bibitem{0301101}
J.~M.~Cline and H.~Firouzjahi,
%``Real-time D-brane condensation,''
arXiv:hep-th/0301101.
%%CITATION = HEP-TH 0301101;%%

\bibitem{0301179}
A.~Ishida and S.~Uehara,
%``Rolling down to D-brane and tachyon matter,''
arXiv:hep-th/0301179.
%%CITATION = HEP-TH 0301179;%%

\bibitem{0303057}
A.~Sen,
%``Dirac-Born-Infeld action on the tachyon kink and vortex,''
arXiv:hep-th/0303057.
%%CITATION = HEP-TH 0303057;%%

\bibitem{0304197}
Ph.~Brax, J.~Mourad and D.~A.Steer, 
arXiv:hep-th/0304197.
%%CITATION = HEP-TH 0304197;%%

\bibitem{0012080}
G.~Arutyunov, S.~Frolov, S.~Theisen and A.~A.~Tseytlin,
%``Tachyon condensation and universality of DBI action,''
JHEP {\bf 0102}, 002 (2001)
[arXiv:hep-th/0012080].
%%CITATION = HEP-TH 0012080;%%

\bibitem{0104218}
N.~D.~Lambert and I.~Sachs,
%``On higher derivative terms in tachyon effective actions,''
JHEP {\bf 0106}, 060 (2001)
[arXiv:hep-th/0104218].
%%CITATION = HEP-TH 0104218;%%

\bibitem{0202079}
K.~Hashimoto and S.~Nagaoka,
%``Realization of brane descent relations in effective theories,''
Phys.\ Rev.\ D {\bf 66}, 026001 (2002)
[arXiv:hep-th/0202079].
%%CITATION = HEP-TH 0202079;%%

\bibitem{0208142}
P.~Mukhopadhyay and A.~Sen,
%``Decay of unstable D-branes with electric field,''
JHEP {\bf 0211}, 047 (2002)
[arXiv:hep-th/0208142].
%%CITATION = HEP-TH 0208142;%%

\bibitem{0304180}
C.~j.~Kim, Y.~b.~Kim and C.~O.~Lee,
%``Tachyon kinks,''
JHEP {\bf 0305} (2003) 020
[arXiv:hep-th/0304180].

\bibitem{0301049}
S.~J.~Rey and S.~Sugimoto,
%``Rolling tachyon with electric and magnetic fields: T-duality 
%approach,''
Phys.\ Rev.\ D {\bf 67}, 086008 (2003)
[arXiv:hep-th/0301049].

\bibitem{0305011}
A.~Sen,
%``Open and closed strings from unstable D-branes,''
arXiv:hep-th/0305011.

\bibitem{0002223}
O.~Bergman, K.~Hori and P.~Yi,
%``Confinement on the brane,''
Nucl.\ Phys.\ B {\bf 580}, 289 (2000)
[arXiv:hep-th/0002223].
%%CITATION = HEP-TH 0002223;%%

\bibitem{0009061}
G.~W.~Gibbons, K.~Hori and P.~Yi,
%``String fluid from unstable D-branes,''
Nucl.\ Phys.\ B {\bf 596}, 136 (2001)
[arXiv:hep-th/0009061].
%%CITATION = HEP-TH 0009061;%%

\bibitem{0010240}
A.~Sen,
%``Fundamental strings in open string theory at the tachyonic vacuum,''
J.\ Math.\ Phys.\  {\bf 42}, 2844 (2001)
[arXiv:hep-th/0010240].
%%CITATION = HEP-TH 0010240;%%

\bibitem{9901159}
P.~Yi,
%``Membranes from five-branes and fundamental strings from Dp branes,''
Nucl.\ Phys.\ B {\bf 550}, 214 (1999)
[arXiv:hep-th/9901159].
%%CITATION = HEP-TH 9901159;%%

\bibitem{0209034}
G.~Gibbons, K.~Hashimoto and P.~Yi,
%``Tachyon condensates, Carrollian contraction of Lorentz group, and  
%fundamental strings,''
JHEP {\bf 0209}, 061 (2002)
[arXiv:hep-th/0209034].
%%CITATION = HEP-TH 0209034;%%

\bibitem{0305092}
C.~Kim, Y.~Kim, O.~K.~Kwon and C.~O.~Lee,
%``Tachyon kinks on unstable Dp-branes,''
arXiv:hep-th/0305092.

\bibitem{LIND}
U.~Lindstrom and R.~von Unge,
%``A picture of D-branes at strong coupling,''
Phys.\ Lett.\  {\bf B403}, 233 (1997)
[hep-th/9704051]; \\
%%CITATION = HEP-TH 9704051;%%
H.~Gustafsson and U.~Lindstrom,
%``A picture of D-branes at strong coupling. II: Spinning partons,''
Phys.\ Lett.\  {\bf B440}, 43 (1998)
[hep-th/9807064]; \\
%%CITATION = HEP-TH 9807064;%%
U.~Lindstrom, M.~Zabzine and A.~Zheltukhin,
%``Limits of the D-brane action,''
JHEP {\bf 9912}, 016 (1999)
[hep-th/9910159].
%%CITATION = HEP-TH 9910159;%%

\bibitem{9507048}
P.~K.~Townsend,
%``P-brane democracy,''
arXiv:hep-th/9507048.

\bibitem{mst}   
R. Dijkgraaf, E. Verlinde, H. Verlinde, 
%"Matrix String Theory", 
Nucl.Phys. B500 (1997) 43-61, 
arXiv:hep-th/9703030.\\
T. Banks and N. Seiberg, 
%"Strings from Matrices", 
Nucl.Phys. B497 (1997) 41-55,
arXiv:hep-th/9702187.\\
R. Dijkgraaf, E. Verlinde and H. Verlinde, 
%"Notes on Matrix and Micro Strings",
Nucl.Phys.Proc.Suppl. 62 (1998) 348-362,
arXiv:hep-th/9709107.\\
L. Motl, 
%"Proposals on nonperturbative superstring interactions", 
arcXiv:hep-th/9701025.\\
R. Schiappa, 
%"Matrix Strings in Weakly Curved Background Fields", 
Nucl.Phys. B608 (2001) 3-50,
arXiv:hep-th0005145.\\
P. J. Silva, 
%"Matrix string theory and the Myers effect,
JHEP 0202 (2002) 004,
arXiv:hep-th/0111121.\\
D.~Brecher, B.~Janssen and Y.~Lozano,
%``Dielectric fundamental strings in matrix string theory,''
Nucl.\ Phys.\ B {\bf 634} (2002) 23
[arXiv:hep-th/0112180].

\bibitem{bfss}  
T. Banks, W. Fischler, S.H. Shenker and L. Susskind, 
%"M Theory As A Matrix Model: A Conjecture, 
Phys.Rev. D55 (1997) 5112-5128, 
arXiv:hep-th/9610043.\\
L. susskind,
%"Another Conjecture about M(atrix) Theory",
arXiv:hep-th/9704080.

\bibitem{9910053}
R.~C.~Myers,
%``Dielectric-branes,''
JHEP {\bf 9912} (1999) 022
[arXiv:hep-th/9910053].
\bibitem{9910052}
W.~I.~Taylor and M.~Van Raamsdonk,
%``Multiple Dp-branes in weak background fields,''
Nucl.\ Phys.\ B {\bf 573} (2000) 703
[arXiv:hep-th/9910052].

\bibitem{0103030}
D.~Mateos and P.~K.~Townsend,
%``Supertubes,''
Phys.\ Rev.\ Lett.\  {\bf 87} (2001) 011602
[arXiv:hep-th/0103030].
\bibitem{0106012}
R.~Emparan, D.~Mateos and P.~K.~Townsend,
%``Supergravity supertubes,''
JHEP {\bf 0107} (2001) 011
[arXiv:hep-th/0106012].
\bibitem{0112054}
D.~Mateos, S.~Ng and P.~K.~Townsend,
%``Tachyons, supertubes and brane/anti-brane systems,''
JHEP {\bf 0203} (2002) 016
[arXiv:hep-th/0112054].
\bibitem{0112034}
D.~s.~Bak and N.~Ohta,
%``Supersymmetric D2 anti-D2 strings,''
Phys.\ Lett.\ B {\bf 527} (2002) 131
[arXiv:hep-th/0112034].
\bibitem{0205265}
D.~s.~Bak, N.~Ohta and M.~M.~Sheikh-Jabbari,
%``Supersymmetric brane anti-brane systems: Matrix model description,  stability and decoupling limits,''
JHEP {\bf 0209} (2002) 048
[arXiv:hep-th/0205265].

\bibitem{0105095}
J.~H.~Cho and P.~Oh,
%``Super D-helix,''
Phys.\ Rev.\ D {\bf 64} (2001) 106010
[arXiv:hep-th/0105095].
%%CITATION = HEP-TH 0105095;%%

\bibitem{0304210}
J.~Gomis, T.~Mateos, P.~J.~Silva and A.~Van Proeyen,
%``Supertubes in reduced holonomy manifolds,''
arXiv:hep-th/0304210.

\end{thebibliography}
\end{document}